\documentclass[preprint,aps,draft]{revtex4}
\usepackage[dvips]{graphicx}

\begin{document}
\title{An external potential dynamic study on the formation of
interface in polydisperse polymer blends}
\author{Shuanhu Qi}
\author{Xinghua Zhang}
\author{Dadong Yan}
\email{yandd@iccas.ac.cn} \affiliation
{Beijing National Laboratory for Molecular Sciences (BNLMS), \\
 Joint Laboratory of Polymer Science and Materials,\\
Institute of Chemistry, Chinese Academy of Sciences, Beijing
100080, China}

\date{\today}

\begin{abstract}
\bigskip
The formation of interface from an initial sharp interface in
polydisperse A/B blends is studied using the external potential
dynamic method. The present model is a nonlocal coupling model as
we take into account the correlation between segments in a single
chain. The correlation is approximately expressed by Debye
function and the diffusion dynamics are based on the Rouse chain
model. The chain length distribution is described by the
continuous Schulz distribution. Our numerical calculation
indicates that the broadening of interface with respect to time
obeys a power law at early times, and the power law indexes are
the same for both monodisperse and polydisperse blend. The power
law index is larger than that in the local coupling model. However
there is not a unified scaling form of the broadening of the
interface width if only the interfacial width at equilibrium is
taken into account as the characteristic length of the system,
because the correlation makes an extra characteristic length in
the system, and the polydispersity is related to this length.
\end{abstract}

\maketitle

\section{INTRODUCTION}

Polymers are often incompatible. When they are mixed together, an
interface between them will occur. Because many properties of the
blends are finally determined by the thickness of the interface
and the concentration profile of the polymer across the interface,
the equilibrium properties and dynamics of the interface attracted
increased attention and gained widespread studies. The broadening
of initially sharp interface between two types of polymers is of
considerable interest.

Self-consistent field theory is a powerful method in the study of
inhomogenous equilibrium systems \cite{Sanchez,Glenn}. It can give
detailed distribution of the monomer concentration across
interface and specify other thermodynamic and structure quantities
of interest. In order to study the kinetic pathway towards its
equilibrium state. A method called ``dynamic mean-field theory"
(DMFT) \cite{Doi,Yeung} or ``dynamic density functional theory"
(DDFT) \cite{Fraaije1,Fraaije2,Maurits} was developed. It is a
generalized time-dependent Ginzburg-Landau theory for conserved
order parameters. The volume fraction (concentration) is often
chosen as the order parameter in polymer blends. In this theory,
the free energy is calculated using the mean-field approximation,
i.e., in each time step, the concentration and the potential
should satisfy the self-consistent equations. This free energy is
more exact than the Flory-Hugins type and the
Ginzberg-Landau-Wilson type free energy which are both a local
free energy plus the square gradient term of the concentration. In
DMFT or DDFT, the Onsager kinetic coefficients are crucial
quantities. Various chain dynamics and interactions will result
different forms of kinetic coefficients. A simple one is the local
coupling form which neglects the non-local interactions. In fact,
sometimes the kinetic coefficients are assumed to be constants. If
the dynamics of polymer chains are described by Rouse model, a
different form of kinetic coefficient will be obtained, and the
coupling between concentrations can be approximately described by
Debye function in a homogenous homopolymer melt \cite{Maurits}. In
the reptation regime, many researchers argued that the forms of
the kinetic coefficient were the same as that for Rouse dynamics
in a homogeneous polymer melt
\cite{Pincus,Binder,Kawasaki1,Kawasaki2}, i.e., proportional to
the Debye function. However, Maurits and Fraaije \cite{Maurits}
found that the kinetic coefficients of the two different dynamics
are not the same though do not differ very much. Wang and Shi
\cite{Wang} studied the interdiffusion process between
incompatible polymers using the Flory-Huggins type free energy
plus the square gradient term of the concentration. The growing of
the interfacial width $W$ obeys the general scaling form
$W(t)\propto t^\alpha$. It is found that the exponent $\alpha$ is
smaller that $0.25$ for all the cases in their studies. As the
square gradient approximation is only valid near the critical
point, it is appropriate to obtain a more exact free energy which
should also be valid far away from the critical point. The DMFT is
the desirable method. Using DMFT with the local coupling
approximation, Yeung and Shi \cite{Yeung} studied the dynamics of
polymer interfaces. They found that the exponent $\alpha$ is about
0.25 at early times, and is independent of the Flory-Huggins
parameter $\chi$ and the chain length. The interfacial width
saturates to its equilibrium thickness at long times. Many
experiments also focused on the interdiffusion process of a system
with an initially sharp interface
\cite{Klein1,Klein2,Klein3,MF,WB}. These experiments found that
the broadening of the interface obeys an power law at early times,
and it needs a very long time to saturate to its equilibrium
thickness. It was also found that the exponent $\alpha$ was in a
range from 0.25 to 0.4, and decreases as the temperature is
lowered \cite{Klein1}. One argued that the quick increase of the
initial broadening is expected to be controlled by the fast
single-chain mobility, while the long time saturation to the
equilibrium thickness is due to the large-scale hydrodynamic flow
\cite{WB}.

The dynamics of polymer interfaces are often described by the time
evolution of the concentration of the chains across the
interfaces. In DMFT, this can be done by solving the
time-dependent Ginzburg-Landau equation which the concentration
satisfies. As this equation is usually solved numerically in real
space, the Onsager coefficient is approximated as a local coupling
form for the simplicity and saving time in calculation
\cite{Doi,Yeung,Fraaije1}. Since the concentration is a conserved
quantity and it is linear rated to the auxiliary potential field
\cite{Schmid,Muller}, we expect that the auxiliary potential field
is also conserved. The time-dependent Ginzburg-Landau equation
with respect to concentration can be transformed to the form in
which the auxiliary potential field is treated as the variable.
Then we can obtain the external potential dynamic (EPD) model, as
the equation is for the auxiliary potential fields. The EPD model
is first proposed by Maurits and Fraaije. There are some
advantages of the EPD method compared to DMFT method
\cite{Reister}. It incorporates a non-local coupling Onsager
coefficient corresponding to the Rouse dynamics. In the spectral
space, this can be realized through a local Onsager coefficient.
Also, it is proved that numerically the EPD equation converges
more easily and faster. In the present work we adopt the EPD
method.

Polymer chains in the real system are always polydisperse, and
polydispersity may play important roles in determining the
properties of materials. Polydispersity enriches the phase
behavior at equilibrium \cite{Sollich} and decreases the free
energy barrier of nucleation in meta-stable state of polymer
blends \cite{Qsh}. The effort of polydispersity on the profile of
the interface at equilibrium are studied by Fredrickson and Sides
\cite{GHF} in polydisperse polymer blends. In the present work, we
studied the dynamics of interfaces in polydisperse blends.

This paper is organized as follows. In Sec. \ref{sec. 2} we
present the derivation of the EPD model based on the Rouse chain
dynamics. The methods for solving the EPD equations are also
talked. In Sec. \ref{sec. 3} the main results and discussion are
presented. In Sec. \ref{sec. 4} we summarizes our conclusions.

\section{External potential dynamic model \label {sec. 2}}
In the present work we consider an incompressible polymer blend of
type A and type B linear, flexible homopolymers in which both
species have polydisperse chain lengths. These two species of
chains are modeled as Gaussian chains. To study the interfacial
problems, it is convenient to work in the canonical ensemble. For
simplicity the distributions of chain length for both species are
described by the same prescribed function $P(N)$, where $N$ is the
degrees of polymerization, thus they have the same number- and
weight-average chain lengths. Also both A and B monomers are
assumed to have the same monomeric volume $\rho_0^{-1}$ and Kuhn
lengths $b$. In this section we first derived the time evolution
equation the concentration satisfies based on the Rouse dynamics
in the polydisperse A/B blend. Similar derivation in the
monodisperse case can be found \cite{Maurits}. Then this equation
of motion has been transformed to the EPD form. The EPD equations
are then solved numerically.

\subsection{Rouse dynamics}

Suppose that a polymer chain of type A with length N is subjected
to an external force $f_A[\mathbf R_N(s)]$, where $\mathbf R_N(s)$
denotes the position of monomer $s$. In the Rouse regime, as the
correlations due to internal forces relax faster than the
coarse-grained collective dynamics. Thus the chain can be
considered as drifting with a constant velocity\cite{Maurits,MDoi}
\begin{equation}\label{drift}
v_d=\frac{D_0}{N}\int_0^Ndsf_A\left[\mathbf R_N(s)\right],
\end{equation}where $D_0$ is the diffusion coefficient of a
monomer. It is important to found from Eq.~(\ref{drift}) that the
drift velocities of chains with different lengths are different,
while the velocities are the same for all beads in the same chain.
The microscopic density of A species is defined as
\begin{equation}\label{density}
\hat\rho_A(\mathbf r)\equiv\sum_{N=1}^{\infty}\sum_{i=1}^{n_{AN}}
\int_0^Nds\delta\left[\mathbf r-\mathbf R_N^i(s)\right].
\end{equation}Here, $n_{AN}$ is the number of A chains for chain
length $N$. The density can be alternatively expressed in the
continuous limit by the following form,
\begin{equation}\label{density2}
\hat\rho_A(\mathbf r)=n_A\int_0^\infty dNP(N)\int_0^N
\delta\left[\mathbf r-\mathbf R_N(s)\right],
\end{equation}where $n_A$ is the total number of A chains.
According to the equation of continuity and using the chain rule,
we obtain,
\begin{equation}
\frac{\partial}{\partial t}\hat\rho_A(\mathbf r,t)=
-n_A\int_0^\infty dNP(N)\int_0^N ds\nabla_{\mathbf r}\delta
\left[\mathbf r-\mathbf R_N(s)\right]\cdot\frac{\partial \mathbf
R_N(s)}{\partial t}.
\end{equation} We then replace $\partial \mathbf R_N(s)/\partial t$ by
Eq.~(\ref{drift}), and the external force can be expressed by
minus the gradient of the chemical potential. By some calculation
we obtain,
\begin{equation}
\frac{\partial}{\partial t}\hat\rho_A(\mathbf
r,t)=D_0n_A\nabla_{\mathbf r}\cdot\int d\mathbf r'\int_0^\infty
dN\frac{P(N)}{N}\int_0^Nds\int_0^Nds'\delta[\mathbf r-\mathbf
R_N(s)] \delta[\mathbf r'-\mathbf R_N(s')]\nabla_{\mathbf
r'}\mu_A(\mathbf r'),
\end{equation}where $\mu_A$ is the chemical potential for A species.
The concentration or the volume fraction of A species is defined
by $\phi_A(\mathbf r)=\langle\hat\rho_A\rangle/\rho_0$, where
$\langle ...\rangle$ denotes the ensemble average. Then we can
obtain
\begin{equation}\label{Debye}
\frac{\partial}{\partial t}\phi_A(\mathbf r,t)=
D_n\overline{\phi}_A\nabla\cdot\int d\mathbf r'\int_0^\infty
dNP(N)Ng_D(\mathbf r-\mathbf r',N)\nabla_{\mathbf r'}\mu_A(\mathbf
r').
\end{equation}Here $D_n=D_0/N_n$, $N_n$ is the number-average
chain length of A species, $\overline{\phi}_A$ is the average
volume fraction, $\overline{\phi}_A=n_AN_n/\rho_0V$, where $V$ is
the volume of the system. The following relation has been used in
deriving Eq.~(\ref{Debye}), $\int_0^Nds\int_0^Nds'\delta[\mathbf
r-\mathbf R_N(s)] \delta[\mathbf r'-\mathbf
R_N(s')]=N^2g_D(\mathbf r-\mathbf r',N)/V$, where $g_D(\mathbf
r-\mathbf r',N)$ is the Debye function with chain length $N$, and
it is valid for Gaussian chains \cite{MDoi}. Similarly we can
obtain the time evolution equation for $\phi_B(\mathbf r,t)$.
However, $\phi_A(\mathbf r,t)$ and $\phi_B(\mathbf r,t)$ are not
independent, they should satisfy the incompressible condition,
$\phi_A(\mathbf r,t)+\phi_B(\mathbf r,t)=1$. The incompressible
condition can be incorporated to the dynamic equation through
introducing a potential $U(\mathbf r)$ which can be added to the
chemical potential. After eliminating this potential from the time
evolution equations for $\phi_A(\mathbf r,t)$ and $\phi_B(\mathbf
r,t)$, we can obtain,
\begin{equation}\label{DMFT}
\frac{\partial}{\partial t}\phi(\mathbf r,t)=\nabla\cdot \int
d\mathbf r'\Lambda(\mathbf r,\mathbf r')\nabla_{\mathbf r'}
\mu_{\phi}(\mathbf r').
\end{equation}Here, $\phi(\mathbf r)
=\phi_A(\mathbf r)-\phi_B(\mathbf r)$ is the concentration
difference, $\Lambda(\mathbf r,\mathbf r')=2D_n\overline{\phi}_A
\overline{\phi}_B\int_0^\infty dNP(N)Ng_D(\mathbf r-\mathbf r',N)$
is the Onsager kinetic coefficient. $\overline{\phi}_B$ is the
average volume fraction of B species, $\overline{\phi}_B=
n_BN_n/\rho_0V$, where $n_B$ is the total number of B chains.
$\mu_{\phi}(\mathbf r)\equiv\mu_A(\mathbf r)-\mu_B(\mathbf r)$ is
the chemical potential difference. In the self-consistent theory,
the chemical potentials are given by $\delta F/\rho_0\delta
\phi_A$ and $\delta F/\rho_0\delta\phi_B$. $F$ is the free energy
in the canonical ensemble:
\begin{equation}
F=\rho_0\int d\mathbf r\chi\phi_A(\mathbf r)\phi_B(\mathbf
r)-\rho_0\sum_{\alpha=A,B}\int d\mathbf r\omega_{\alpha}(\mathbf
r)\phi_\alpha(\mathbf r)-\sum_{\alpha=A,B}n_\alpha\int_0^\infty
dNP(N)\ln Q_\alpha(\omega_\alpha,N).
\end{equation}Here, $\omega_\alpha$ are the auxiliary fields
conjugated to $\phi_\alpha$, $Q_\alpha[\omega_\alpha,N]$ are the
single chain partition functions of the chain length $N$ for
$\alpha$ species.

\subsection{External potential dynamic equation}
We have mentioned that concentration is a conserved quantity and
it is linear rated to the auxiliary potential field, then the
external potential field can be treated as a conserved variable.
The dynamics of this potential can be described by the equation of
the time-dependent Ginzburg-Landau form. The free energy of the
system can also be written as a functional of the external
potentials \cite{Glenn,GHF,Ceniceros}, which has the form of
\begin{eqnarray}
H[\mu_+,\mu_-] &=& \int d\mathbf r\left(\frac{\rho_0}{\chi}
\mu_-^2-\rho_0\mu_+\right)-n_A\int_0^\infty dNP(N)\ln
Q_A[\mu_+-\mu_-,N] \nonumber \\ &-&n_B\int_0^\infty dNP(N)\ln
Q_B[\mu_++\mu_-,N].
\end{eqnarray}Here, $\mu_+\equiv(\omega_A+\omega_B)/2$,
$\mu_-\equiv(\omega_B-\omega_A)/2$ are treated as the external
potentials. The external field $\mu_-$ couples the the
concentration difference $\phi_A-\phi_B$, while $\mu_+$ couples to
the total concentration $\phi_A+\phi_B$. As there are two external
fields in $H$, the numerical evaluation of the functional integral
on the fields is difficult, often a saddle point approximation
will be made when the fields $\mu_+$ is integrated \cite{Muller},
which means $\left.\delta H[\mu_+,\mu_-]/\delta \mu_+\right|
_{\mu_+=\mu_+^*}=0$. This is consistent with the incompressible
condition, i.e.
\begin{equation}\label{Incompressible}
\phi_A[\mu_+^*-\mu_-]+\phi_B[\mu_+^*+\mu_-]=1.
\end{equation}
The external potential dynamics can be expressed as
\begin{equation}
\frac{\partial}{\partial t}\mu_-(\mathbf r)=\nabla\cdot\int
d\mathbf r' \Lambda_{\mbox{\scriptsize EPD}}(\mathbf r,\mathbf
r')\nabla_{\mathbf r'}\frac{\delta H[\mu_+^*,\mu_-]}
{\rho_0\delta\mu_- }.
\end{equation}Here, $\Lambda_{\mbox{\scriptsize EPD}}
(\mathbf r,\mathbf r')$ is the Onsager coefficient in EPD model,
and it is related to $\Lambda(\mathbf r,\mathbf r')$.  The
explicit expression of $\Lambda(\mathbf r,\mathbf r')$ can be
obtained from Eq.~(\ref{DMFT}). According to the chain rule
\begin{equation}\label{transform}
\frac{\partial \phi(\mathbf r,t)}{\partial t}=\int d\mathbf r'
\frac{\delta \phi(\mathbf r,t)}{\delta\mu_-(\mathbf r',t)}
\frac{\partial\mu_-(\mathbf r',t)}{\partial t},
\end{equation}where $\partial\mu_+/\partial t=0$ has been used.
According to the linear response theory, using the saddle point
approximation, for Gaussian chains at any time we can obtain
\begin{equation}\label{phi}
\phi_A(\mathbf r)=-\frac{n_A}{\rho_0}\int_0^\infty
dNP(N)\frac{\delta Q_A[\omega_A]}{\delta\omega_A(\mathbf r)},
\end{equation}and
\begin{equation}
\frac{\delta \phi_A(\mathbf r)}{\delta\omega_A(\mathbf r')}=-
\frac{n_A}{\rho_0}\int_0^\infty dNP(N)\frac{\delta^2 \ln
Q_A}{\delta\omega_A(\mathbf r)\delta\omega_A(\mathbf r')}=-
\frac{n_A}{\rho_0V}\int_0^\infty dNP(N)N^2g_D(\mathbf r-\mathbf
r',N).
\end{equation}Similarly, we can obtain the expressions for
$\delta \phi_B(\mathbf r)/\delta\omega_B(\mathbf r')$, and $\delta
\phi_A(\mathbf r)/\delta\omega_B(\mathbf r')=\delta \phi_B(\mathbf
r)/\delta\omega_A(\mathbf r')=0$. For the incompressible Gaussian
chain system, after some calculations, we can get
\begin{equation}\label{linear}
\frac{\delta\phi(\mathbf r)}{\delta\mu_-(\mathbf r')}=-2
\frac{\delta[\phi_A(\mathbf r)-\phi_B(\mathbf r)]}
{\delta[\omega_A(\mathbf r')-\omega_B(\mathbf r')]}=4
\overline{\phi}_A\overline{\phi}_B\int_0^\infty dNP(N)
\frac{N^2}{N_n}g_D(\mathbf r-\mathbf r',N).
\end{equation}Insert Eq.~(\ref{linear}) into
Eq.~(\ref{transform}) and compare with Eq.~(\ref{DMFT}), and also
through a Fourier transformation, the EPD equation can be obtained
in the spectral space
\begin{equation}\label{EPD}
\frac{\partial \mu_-(\mathbf q)}{\partial t}=\frac{D_n\chi}{2}
\frac{\int_0^\infty dNP(N)Ng_D(\mathbf q,N)} {\int_0^\infty
dNP(N)\frac{N^2}{N_n}g_D(\mathbf q,N)}q^2
\left[-\frac{2\mu_-(\mathbf q)} {\chi}+\phi_A(\mathbf
q)-\phi_B(\mathbf q)\right],
\end{equation}where $\phi_A$ and $\phi_B$ are functionals of
$\mu_+^*$ and $\mu_-$, and $\mu_+^*$ is determined by the
incompressible condition, i.e., Eq.~(\ref{Incompressible}). The
Onsager coefficient in spectral space has the form of
\begin{equation}
\Lambda_{\mbox{\scriptsize EPD}}(\mathbf q)=\frac{D_n\chi}{2}
\frac{\int_0^\infty dNP(N)Ng_D(\mathbf q,N)} {\int_0^\infty
dNP(N)\frac{N^2}{N_n}g_D(\mathbf q,N)},
\end{equation}where $g_D(\mathbf q,N)=2(x+e^{-x}-1)/x^2$,
$x=R_g^2q^2$, and $R_g=Nb^2/6$ is the radius of gyration. In the
monodisperse limit, $P(N)=\delta(N-N_n)$, then
$\Lambda_{\mbox{\scriptsize EPD}}(\mathbf q)=D_0\chi/2N_n$, which
is consistent with the result obtained by M\"{u}ller and Schmid
\cite{Muller}. In the present work, we adopt the continuous Schulz
distribution for reflecting realistic chain length distribution.
However, it is not intrinsic to our topics. The Schulz
distribution has the form of
\begin{equation}
P(N)=\frac{N^{k-1}(k+1)^k}{N_w^{\,k}\Gamma(k)}
\mbox{exp}[-(k+1)N/N_w],
\end{equation}where $N_w$ is the weight-average chain
length, $k$ is a parameter related to the polydispersity index, a
smaller $k$ corresponds to a more polydisperse distribution, and
the infinity of $k$ corresponds to the monodisperse case.

\subsection{Numerical calculation}
In the present work, we consider a one-dimensional system, and
$\overline{\phi}_A=\overline{\phi}_B=1/2$. The EPD equation
(Eq.~(\ref{EPD})) and Eq.~(\ref{Incompressible}) are closed and
solved numerically. We adopt the numerical method developed by
Ceniceros and Fredrickson \cite{Ceniceros,Glenn}, which is a
semi-implicit Seidel relaxation scheme. In the present work we
studied the healing process of an initially sharp interface. The
initially sharp interface was prepared by a tangent function, and
for A species, it is
\begin{equation}
\phi_{AI}(x)=\frac{\phi_{Ap}+\phi_{Am}}{2} +\frac{\phi_{A
p}-\phi_{Am}}{2\tanh\eta}\tanh\left[\eta\cos\left(\frac{2\pi x}
{l_x}\right)\right],
\end{equation}where $\phi_{Ap}$ and $\phi_{Am}$ are the
concentrations at the boundaries in a phase-separated system at
equilibrium, $p$ denotes the A-rich boundary, while $m$ denotes
the A-poor boundary. $l_x$ is the length of the system, $x$ ranges
from $-l_x/2$ to $l_x/2$. The parameter $\eta$ determines the
sharpness of the interface, and a larger $\eta$ corresponds to a
sharper interface. In the present work, we choose $\eta=100$ which
is large enough in our study. The initial concentration for B
species is $\phi_{BI}(x)=1-\phi_{AI}(x)$. However, in order to
proceed the evolution of the EPD equation, we need the initial
external potentials $\mu_{I-}(x)$ and $\mu_{I+}(x)$ which target
the initial concentrations. This is realized by numerically
solving the following equations
\begin{eqnarray}
\phi_A[\mu_{I+}(x)-\mu_{I-}(x)]-\phi_{AI}(x)=0, \nonumber\\
\phi_B[\mu_{I+}(x)+\mu_{I-}(x)]-\phi_{BI}(x)=0.
\end{eqnarray}We also use the semi-implicit Seidel relaxation
scheme.

In the calculation of the concentrations $\phi_\alpha(\mathbf r)$
($\alpha=$A,B), the infinite integrations are performed using a
Gauss-Laguerre quadrature formula. The concentration (see
Eq.~(\ref{phi})) can be expressed as
\begin{equation}
\phi_\alpha(x)=\frac{(k+1)\overline{\phi}_\alpha}
{\Gamma(k+1)}\int_0^\infty dMe^{-M}\frac{M^{k-1}}
{Q_\alpha[M/k+1,\omega_\alpha]}\int_0^{M/k+1}dsq_\alpha
\left(x,\frac{M}{k+1}-s\right)q_\alpha(x,s),
\end{equation}where $M=(k+1)N/N_w$. The Gauss-Laguerre quadrature
formula is
\begin{equation}
\int_0^\infty dMe^{-M}f(M)= \sum_{i=1}^{n_{_G}}\lambda_if(M_i),
\end{equation}where the abscissas ($M_i$) and weights
($\lambda_i$) can be find in a mathematical handbook. This formula
converges very rapidly, and we find that 8 points are sufficiently
accurate for all cases. In the present work, we adopt $n_{_G}=8$.

The relations between the partition functions of the single chain
($Q_\alpha$) and the end-integrated propagators ($q_\alpha$) are
$Q_\alpha[N,\omega_\alpha]=\int dxq_\alpha(x,N)/l_x$, where
$q_\alpha$ satisfy the modified diffusion equations, which have
the form of
\begin{equation}
\frac{\partial q_\alpha(x,s)}{\partial s}=\frac{b^2}{6}
\frac{\partial^2q_\alpha(x,s)}{\partial x^2}-\omega_\alpha(x)
q_\alpha(x,s),
\end{equation}with the initial conditions $q_\alpha(x,0)=1$.
The modified diffusion equations are solved numerically using the
pseudo-spectral method \cite{Glenn,GHF} with periodic boundary
conditions (This results two A-B interfaces). This method is
unconditionally stable in any number of space dimension and have
higher accuracy \cite{Qiang} than the Crank-Nicholson
semi-implicit schemes in our experience. For the convenience some
quantities are scaled in the calculation
\begin{equation}
x\rightarrow x/R_{gw}, \ s\rightarrow s/N_w, \ t\rightarrow
tD_n/R_{gw}^2,
\end{equation}where $R_{gw}=N_wb^2/6$. We also used $\chi N_w$
instead of $\chi$ to characterize the incompatibility. The length
of the system were chosen as $l_x=40$, and the number of spatial
grid was $N_x=512$, which means a mesh size of $\Delta
x=0.078125$. In solving the modified diffusion equations the
prescribed contour steps were $\triangle s\leq 0.0025$, which
ensured a eight-figure accuracy of the concentration profile. The
time step was chosen smaller at early time and larger approaching
equilibrium at later times. We also conformed that our results
were independent of the mesh size and the system size by varying
$\Delta x$ and $l_x$.

\section{results and discussion \label {sec. 3}}
In order to study the effect of polydispersity on the formation of
interface, we first investigated the evolution of interface in the
case of monodisperse A/B blend. The model we used in the
monodisperse case was a simplified one of the present polydisperse
model, which is also a nonlocal coupling one.

Figure 1 shows an example of the time evolution of the density
profile in the monodisperse case, in which the volume fraction of
A species $\phi_A(x)$ at different times for $\chi N$=2.5 is
presented. The initial interface is very sharp, and it evolves to
the equilibrium one eventually at long times. In this process, as
show in Fig.~1 the $\phi$-enhanced and -depleted bumps appear,
which were also observed in the local coupling model \cite{Yeung}.
This can be understood by symmetry as explained by Yeung and Shi
\cite{Yeung}. If we consider the interface at $x=-10$, and the
locations at $x=-20$ and $x=0$ correspond to the phase boundaries
at infinitely far away. The chemical potential is expressed as
$\mu=\delta F/\delta\phi$ ($\phi$ means $\phi_A$), and it equals
to zero at $x=-20$ and $x=0$; also the chemical potential is an
odd function with respect to $x=-10$, hence $\mu=0$ at $x=-10$.
There must be extremum values of the chemical potential between
the interface and the phase boundaries, which is a maximum value
on the A-rich side of the interface and a minimum value on the
A-poor side. The diffusion of the order parameter satisfies
$j_x=-M(\phi)d\mu/d\phi$, where $j_x$ is the current and
$M(\phi)>0$ is a dynamical coefficient. Near the interface, $\phi$
is transported from the $\phi$-rich side of the interface to the
$\phi$-poor side. However, the current changes its sign where the
chemical potential is extremized. Therefore at some place far away
from the interface, $\phi$ must be transported away from the
interface on $\phi$-rich side and towards the interface on
$\phi$-poor side. This is why an enhancement of $\phi$ above the
equilibrium value on the A-rich side and a depletion on the A-poor
side appear in the calculation. The calculation indicates that
bumps are larger at smaller $\chi N$, while they are smaller at
larger $\chi N$, and in the course of time the bumps move to the
phase boundaries.

What we are interested in is the growth law of interfacial width
with respect to time. The density across the interface is not
homogeneous, thus the definition of the interfacial width is not
unique. As a quantitative measure of the interfacial width, we
followed Yeung and Shi, and also Steiner et al. \cite{Klein1} and
chose the inverse of the maximum slope of $\phi_A$ at the
interface
\begin{equation}\label{Width}
W(t)=\left(\left.\frac{1}{\Delta\phi_A}\frac{\partial\phi_A}{\partial
x}\right|_{x=-10}\right)^{-1}.
\end{equation}Here, $\Delta\phi_A$ is the difference between the
bulk value of $\phi_A$ in the two equilibrium phases. It is
obvious that this definition is sensitive to the local structure
of the interface. Other definition, e.g. taking into account the
entire structure of the interface, can alternatively be chosen,
however, it dose not change the scaling relations.

Figure 2 shows the interfacial width defined in Eq.~(\ref{Width})
as a function of time for different $\chi N$. The interfacial
widthes are divided by their equilibrium ones $\xi$. From the
figure it can be seen that only at the very beginning the
broadening of the interface obeys a power law with respect to
time. In order to find this relation we rescale the time variable,
which means a shift of the profile along $t$-axis. As shown in
Fig.~(3), if $W(t)$ is sclaed by $\xi$ and the time also scaled by
$\xi$, all the data from Fig. 2 approximately collapse onto one
single master curve. It means that $W(t)$ satisfies the following
equation
\begin{equation}\label{Widthm}
W(t)=\xi f_{\mbox{\scriptsize mono}}(t/\xi^\beta),
\end{equation}where $f_{\mbox{\scriptsize mono}}$ is a
dimensionless function with $f_{\mbox{\scriptsize
mono}}(\infty)=1$ and $\beta=1.0$. It can be found from the figure
that at the early times, $W(t)$ satisfies the following equation
$W(t)\propto ct^\alpha$, where $c$, a time independent constant,
is related to $\xi$, and $\alpha$ is the power law index. From the
data fitting we found that $\alpha$ approximately equals to 0.38.
A direct scaling analysis using an expansion of the free energy to
the second-order in $\nabla\phi$ by the Cahn-Hilliard dynamics
gives $t^{-1}\propto W^{-4}$ or $W\propto t^{1/4}$. An more exact
form of the free energy, the self-consistent mean field free
energy with the Cahn-Hilliard dynamics also demonstrated $W\propto
t^{1/4}$. It is easy to understand this relation since there is
only one characteristic length $\xi$ the equilibrium width of the
interface in the approximated Cahn-Hillard dynamic system. In the
present model, we take into account the nonlocal coupling between
segments of a single chain, hence, there exists another
characteristic length, the correlation length. It is because of
this correlation length, a simple scaling analysis with one
characteristic length can not predict an exponent of 0.38. An
prefactor of $c$ is also a reflection of the existence of another
characteristic length. This will be more precise in the
polydisperse case, which will be talked later. An exponent of 0.38
lager than 0.25 means that the correlation between segments
quicken the broadening of the interface.

In the following we consider the formation of interface in a
polydisperse A/B blend. In the process of the evolution of the
volume fraction of $A$ species from its initially sharp
distribution to the equilibrium diffusive distribution, the
enhancement of $\phi_A$ above the equilibrium value on the A-rich
side of the interface and a depletion on the A-poor side are also
observed in the numerical calculation. The magnitude of these
bumps becomes smaller when $\chi Nw$ becomes larger, and also the
bumps move to the phase boundary as $t$ grows. It is similar to
the case that in monodisperse blend, and also can be understood by
symmetry.

Figure 4 shows the evolution of interface with respect to time in
the polydisperse polymer blends, the interfacial width is divided
by $\xi$. The polydisperse parameters are chosen as $k=1$, 2, and
3, corresponding to the case that the polydispersity is reducing.
The monodisperse case corresponds to an infinity of $k$. The
incompatibility parameters are chosen as $\chi Nw=3.5, 4, 6, 8$
from weak segregation region to strong segregation region. From
the figure it is seen that only at very early time there is a
power law between the interfacial width and time. In order to find
this relation we also rescale the time variable. However, in the
simulation we found that different forms of scaling of time were
needed for different polydispersities. This is as expected since
the extra characteristic lengths, the correlation lengths are
different in blends with different polydispersities. There is not
a unified scaling expression which contains only one
characteristic length. Figure 5 shows the detailed information of
forms of time scaling under which the data approximately collapse
onto one single master cure for different polydispersities,
respectively. From Fig.~5 (a) for the case of $k=1$ it can be
found that if $W/\xi$ is plotted as function of $t/\xi^{1.9}$, all
data for different incompatibilities approximately collapse onto
one single line; from Fig.~5 (b) and Fig.~5 (c), one can see that
the scaling forms of time are $t/\xi^{1.6}$ and $t/\xi^{1.5}$ for
$k=2$ and $k=3$, respectively. It means that for a polydisperse
polymer blend the broadening of interface satisfies
\begin{equation}\label{Widthp}
W(t)=\xi f_k(t/\xi^\beta),
\end{equation}where $f_k$ is a dimensionless function with
$f_k(\infty)=1$ and $k$ denotes the polydispersity. From the
simulation it is obvious that the parameter $\beta$ decreases as
the polydispersity decreases, and it approaches the value of
monodisperse case 1.0 when $k$ goes to infinity. Equation
(\ref{Widthp}) demonstrates that there is not a unified scaling
form of the broadening of the interface if we only take into
account the equilibrium width of interface as the characteristic
length. An unified scaling form which can include different case
of polydispersities may be constructed if two characteristic
lengths, the equilibrium width of interface and the correlation
length, are taken into account. However, it is out of the present
work. At the early time for all the polydisperse case the
broadening of interface with respect to time satisfies
$W(t)\propto t^\alpha$, where the power law $\alpha\simeq0.38$
which is the same as that in the monodisperse case. Thus we can
conclude that the nonlocal coupling quickens the diffusion of
polymer chains across the interface, while the polydispersity has
no effect on the power law index.

\section{conclusion \label {sec. 4}}
The formation of interface from an initially sharp one in
polydisperse A/B blend is studied using external potential dynamic
method. The chains of Both A and B species are polydisperse, and
their chain length distributions are described by the continuous
Schulz distribution. Based on the Rouse chain model, we derived
the time evolution equation the concentration satisfies. In the
Rouse dynamcis, the drift velocities of chains with different
lengths are different, while the velocities are assumed to be the
same for all beads in the same chain. The collective Rouse
dynamics model is a nonlocal coupling model. This coupling is
reflected by the Onsager coefficient which is related to the Debye
function. The time evolution of the concentration then transformed
to the EPD form, because of which are numerically more efficient
to solve. The EPD equations are solved numerically using a
semi-implicit Seidel relaxation scheme. Our calculation indicates
that the broadening of interfacial width with respect to time
obeys a power law at early times, and the power law indexes are
the same for both monodisperse and polydisperse blend. This means
that the polydispersity does not affect the form of this power
law. The power law index is larger than that in the local coupling
model. It is obvious that the nonlocal coupling quicken the
diffusion of polymer chains across the interface. However there is
not a unified scaling form of the broadening of the interfacial
width if only the equilibrium interfacial width is taken into
account as the characteristic length of the system, because there
exists another characteristic length, the correlation length. The
correlation lengths are different in blends with different
polydispersities. An unified scaling form which can include
different case of polydispersities may be constructed if two
characteristic lengths, the equilibrium width of interface and the
correlation length, are taken into account. This will be studied
in the future.

\newpage
\begin{center}
\textbf{ACKNOWLEDGMENTS}
\end{center}

We thank Prof. An-Chang Shi for the helpful discussion and
suggestions. This work is supported by National Natural Science
Foundation of China (NSFC) 20574085 and the Grant from Chinese
Academy of Sciences KJCX2-YW-206.

\newpage
\clearpage
\begin{figure}
\centerline{\includegraphics[angle=0,scale=1.2,draft=false]{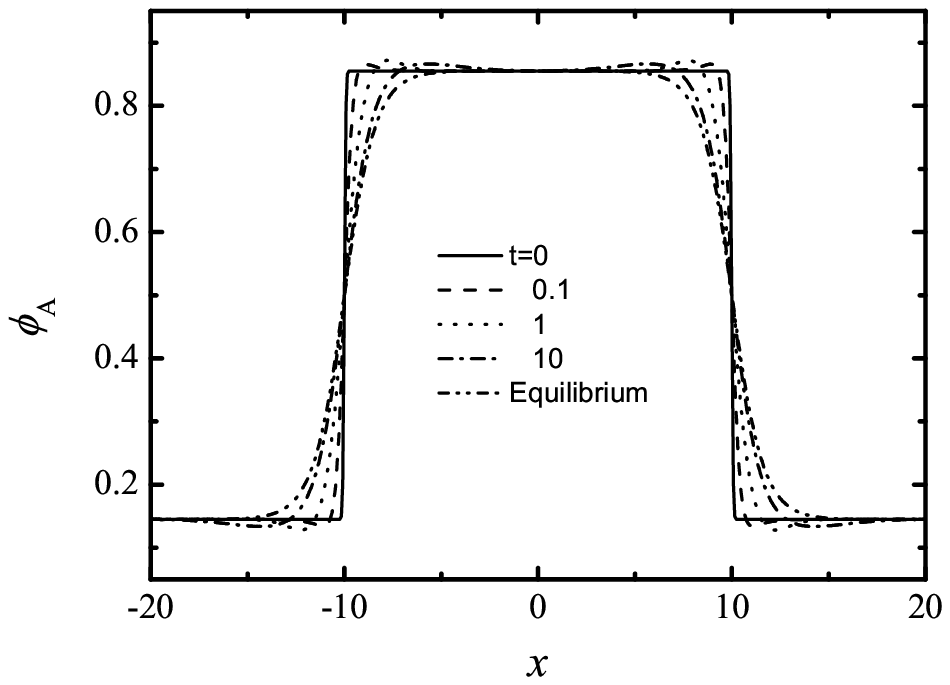}}
\caption{Time evolution of the density profile in the monodisperse
case for $\chi N=2.5$. The $\phi_A$-enhanced and -depleted bumps
appear, which can be explained by symmetry.} \label{01}
\end{figure}

\newpage
\clearpage
\begin{figure}
\centerline{\includegraphics[angle=0,scale=1.2,draft=false]{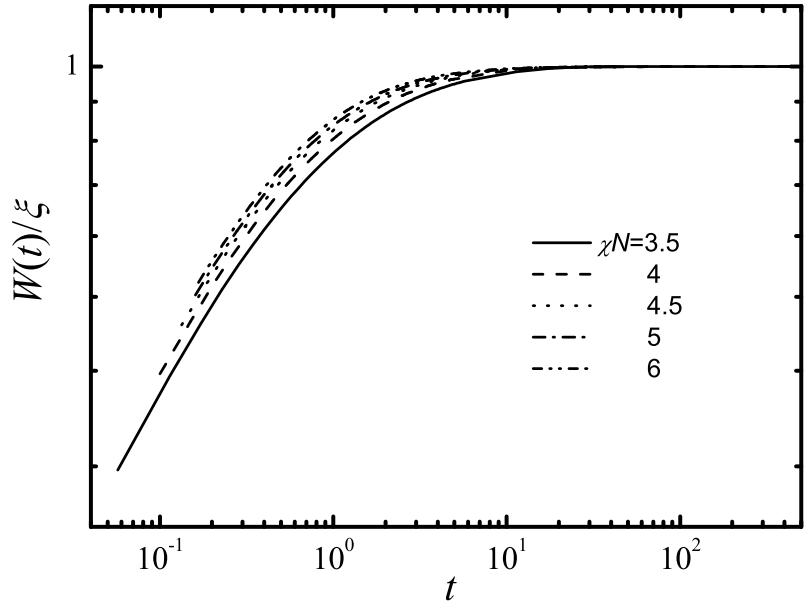}}
\caption{The broadening of interface with respect to time for
different $\chi N$ in the monodisperse case. The interfacial width
is scaled by $\xi$. It shows that at early times the interfacial
widthes broaden fast, and they saturate to the equilibrium ones at
very long times. } \label{01}
\end{figure}

\newpage
\clearpage
\begin{figure}
\centerline{\includegraphics[angle=0,scale=1.2,draft=false]{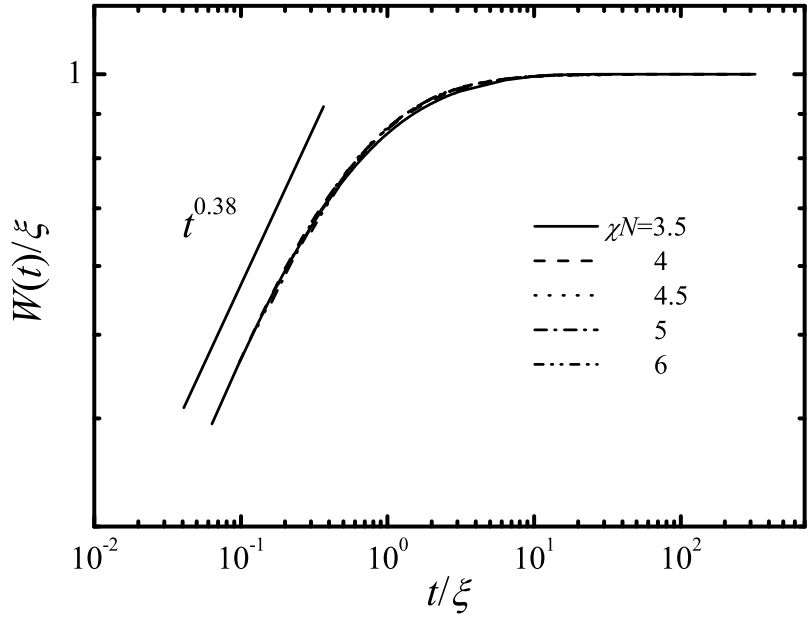}}
\caption{Data from Fig.~2 plotted in a scaled form. Here the
interfacial width is scaled by $\xi$, the time is also scaled by
$\xi$. The data approximately collapse onto a single master curve.
At early times, the broadening of interfacial width obeys a power
law, and the power law index is about 0.38.} \label{01}
\end{figure}

\newpage
\clearpage
\begin{figure}
\centerline{\includegraphics[angle=0,scale=1.2,draft=false]{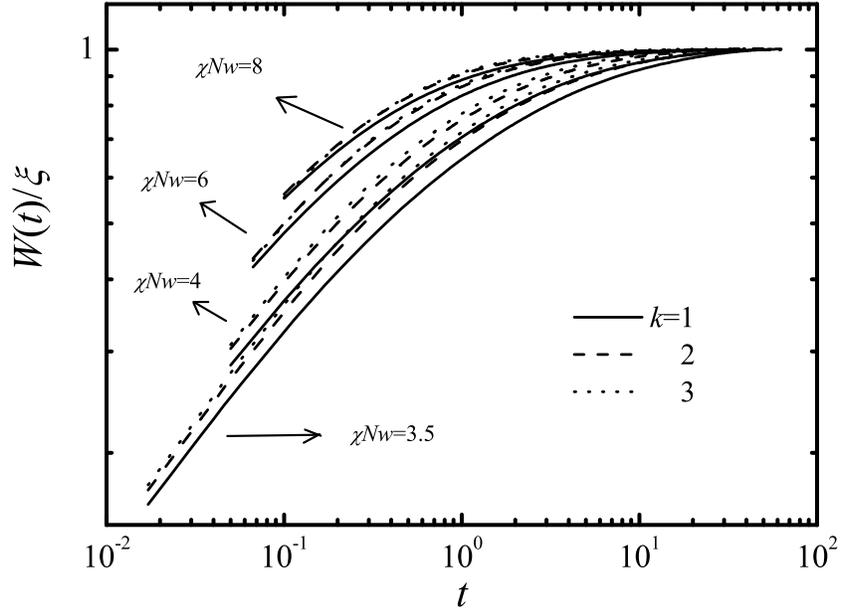}}
\caption{The evolution of interface with respect to time in the
polydisperse case for different $\chi N_w$ and different
polydispersities. The interfacial width is scaled by $\xi$}
\label{01}
\end{figure}

\newpage
\clearpage
\begin{figure}
\centerline{\includegraphics[angle=0,scale=2,draft=false]{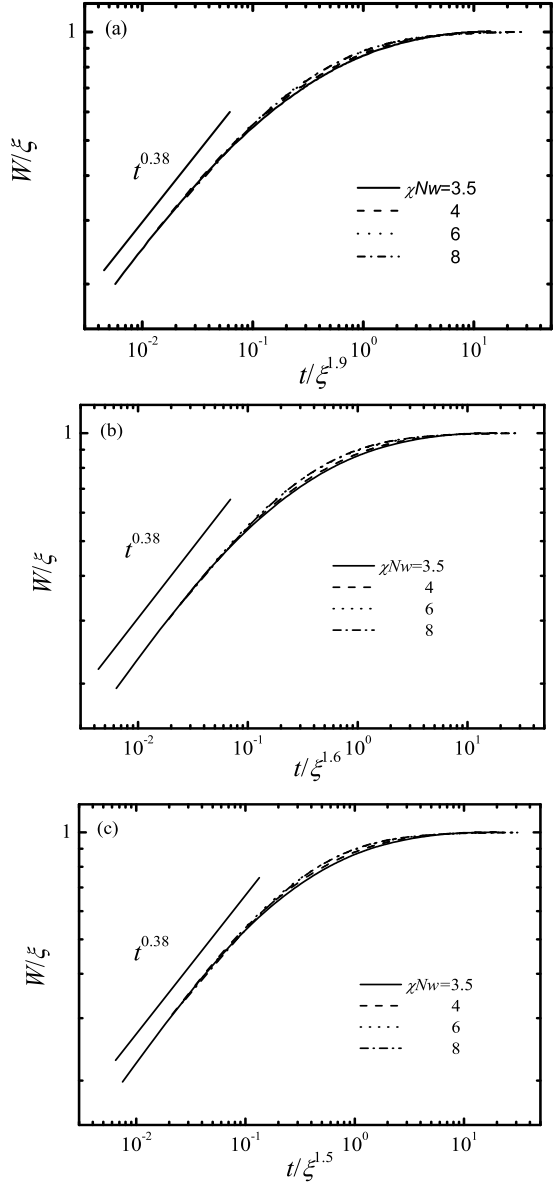}}
\caption{Data from Fig.~3 plotted in a scaled form. Here the
interfacial width is scaled by $\xi$, the time is also scaled by
$\xi^{1.9}$ for $k=1$ (a), by $\xi^{1.6}$ for $k=2$ (b), and by
$\xi^{1.5}$ for $k=3$ (c). The exponent of $\xi$ approaches the
value that in the monodisperse case as $k$ goes to infinity. The
data approximately collapse onto a single master curve for
different polydisperse case, respectively. At early times, the
broadening of interfacial width obeys a power law, and the power
law index is about 0.38 which is independent of the
polydispersity.} \label{01}
\end{figure}
\end{document}